\documentclass[10pt]{iopart}
\usepackage{epsfig,graphicx}
\usepackage{epsfig}
\newcommand{\argmax}{{\rm arg}\max}

\newcommand{\bra}{\langle}
\newcommand{\ket}{\rangle}

\newcommand{\order}{{\mathcal O}}

\newcommand{\sgn}{\textrm{sgn}}

\newcommand{\one}{{\rm 1\!\!I}}

\newcommand{\bnull}{{\mbox{\boldmath $0$}}}
\newcommand{\be}{\begin{equation}}
\newcommand{\ee}{\end{equation}}

\newcommand{\room}{\rule[-0.1cm]{0cm}{0.6cm}}

\newcommand{\R}{{\rm I\!R}}

\newcommand{\bR}{\ensuremath{\mathbf{R}}}

\newcommand{\bpsi}{{\mbox{\boldmath $\psi$}}}

\newcommand{\bvarphi}{{\mbox{\boldmath $\varphi$}}}

\newcommand{\here}{\makebox(0,0)}

\begin{document}

\title{Market response to external events and interventions in spherical minority games}

\author{P Papadopoulos and ACC Coolen}
\address{Department of Mathematics, King's College London\\ The Strand, London WC2R 2LS, UK}
\ead{panagiotis.2.papadopoulos@kcl.ac.uk, ton.coolen@kcl.ac.uk}

\begin{abstract}
We solve the dynamics of large spherical Minority Games (MG) in the presence of non-negligible time dependent external contributions to the overall market bid. The latter represent the actions of market regulators, or other major natural or political events that impact on the market.
In contrast to non-spherical MGs, the spherical formulation allows one to derive closed dynamical order parameter equations in explicit form and work out the market's response to such events fully analytically. We focus on a comparison between the response to stationary versus oscillating market interventions, and reveal profound and partially unexpected differences in terms of transition lines and the volatility.
\end{abstract}

\section{Introduction}

Minority Games (MG)  \cite{1,2} are simple mathematical
models designed to elucidate and explain the origin of the nontrivial macroscopic
fluctuation phenomenology observed in real markets, on the basis of so-called inductive decision making by large numbers of
interacting agents \cite{Brian Arthur}. Their great strength is that they can be solved analytically using methods from the statistical mechanics of disordered systems, in particular with generating functional analysis techniques \cite{DeDom,Ton1,Ton2}.  We refer to the recent textbooks
 \cite{Book1, Book2} for historical backgrounds, the connection between MGs and real markets,  details on mathematical methods, and full references. Now that the standard
 MGs have been solved and understood satisfactorily, attention must turn to generalizing the mathematical technology developed so as to
 apply to models that are more realistic economically.
 In this paper we try to contribute to this aim by studying analytically the dynamical response of MG markets to non-negligible events in the outside world that impact on the
 overall market bid (which in MGs is a proxy for the asset price). These events could be accidental (e.g. natural disasters, changes in resource availability), political (e.g. election results, major management incompetence or corruption scandals),
 or interventions by market regulators. Such ingredients are incorporated easily into the fabric of MG-type models, by simply adding to the overall market bid time-dependent external terms; the tricky stage is to work out mathematically the consequences of such terms. The simplest case
 is that where the external bid term is stationary. Here only minor modifications of the standard formalism are required; see e.g. \cite{ChalletMarsiliOttino} (using the replica method) or \cite{DeSanctisGalla} (using generating functional analysis). For non-stationary external bid contributions, in contrast, one is generally forced to either resort to numerical simulations (see \cite{DeSanctisGalla}), or turn to those generating functional analysis versions that involve explicit representations of the overall bid process, as developed for MGs with real market histories \cite{RealHistories,InnerProduct}. In the latter
 studies only infinitesimal external bid perturbations have been considered so far,  in view of the complexity of the formalism. There is, however, one special class of MG versions where adding time-dependent external bids does not lead to serious mathematical complications:
 the spherical MGs as introduced in \cite{1st Spherical,2nd Spherical}. We show in this paper that here it is still possible to derive fully explicit and exact equations for time dependent order parameters (correlation and response functions); time-dependent external bids are found to be mathematically
 harmless, and one can even allow for (partial)   market
impact correction as in \cite{1st impact}. We focus on comparing the long-time solution of these equations for stationary external bids to those found for oscillating external bids. As expected, these two cases are found to generate very different macroscopic consequences, in terms of phase diagrams and in terms of the volatility. Some of these are intuitively clear, such as the profound impact of oscillating market interventions on the volatility, but some are not at all intuitive, such as the independence of the phase diagram in the case of oscillating external bids on the amplitude of the intervention (in contrast to stationary bids). All our theoretical results are tested against numerical simulations, and find excellent confirmation.

\section{Definitions}

In MGs one considers $N$ agents, labeled usually by
Roman indices $i=1,\ldots,N$. At each time step
$t\in\left\{0,1,2,\ldots\right\}$ of the game each agent $i$
submits a trading action, a `bid' $b_i(t)$, in response to
public information $\mu(t)$ which in fake history MG versions is chosen randomly
and independently  from the set $\{1,\ldots,p\}$ (where $p=\alpha
N$ with $\alpha$ finite). The rescaled total market bid in the game at time $t$ is subsequently defined as
 \begin{eqnarray}
A(t)=  A_e(t)+\frac{1}{\sqrt N}\sum_ib_i(t)
\end{eqnarray}
Here $A_e(t)$ could
represent e.g. random market perturbations, actions by market
regulators, or other external events (natural, social, political, etc.) that
can change the overall asset demand in the market directly. Each agent $i$ has $S$ `look-up table' strategies
$\bR^{ia}=(R_{1}^{ia},\ldots,R_p^{ia})\in\{-1,1\}^p$, with $a=1,\ldots,S$. If agent $i$ decides to use strategy
$a$ at time $t$ in the game, his bid at that stage will be
$b_i(t)=R^{ia}_{\mu(t)}$. In the MG
an agent $i$ finds himself winning at time $t$ if his decision $b_i(t)$ turns out to be opposite in sign to the total bid,
i.e. if $b_i(t)A(t)<0$. All agents monitor the performance of their
strategies, in order to decide which one to use. To this end,
they assign points to each of their strategies based on the
 update rule
 \begin{eqnarray}
p_{ia}(t+1)=p_{ia}(t)-\eta R^{ia}_{\mu(t)}\Big[A(t)-\frac{\kappa}{\sqrt
N}\Big(R^{ia_i(t)}_{\mu(t)}-R^{ia}_{\mu(t)}\Big)\Big]
\label{eq:valuations}
\end{eqnarray}
where $\eta$ is a learning rate (which sets the unit time scale),
and where the first minus sign implements reward for minority decisions. The term
proportional to $\kappa\in[0,1]$ represents a (partial) correction by individual agents of $A(t)$ for their own contribution,
as in \cite{1st impact}. The strategy played by agent $i$ at time $t$ is then $a_i(t)=\argmax_{a\in\{1,\ldots,S\}}p_{ia}(t)$.
In this paper we limit ourselves to $S=2$; viz. two strategies per agent.
It is now sufficient for agents to keep track only of the
differences $q_{i}(t)=\frac{1}{2}[p_{i1}(t)-p_{i2}(t)]$. Upon also replacing in the right-hand side of (\ref{eq:valuations}) the random $\mu(t)$ at each step $t$ by an average over all possible values (the so-called batch version of the game), viz. ${\rm RHS}[\mu(t)]\to p^{-1}\sum_{\mu\leq p}{\rm RHS}[\mu]$, and with $\eta=2\alpha\sqrt{N}$\footnote{The $\sqrt{N}$ ensures that the relevant time scales are $\order(N^0)$; the factor $2\alpha$ leads to simple equations.}
one finds
\begin{eqnarray}
q_{i}(t+1)&=&q_{i}(t)+\theta_{i}(t)-\frac{2}{\sqrt N}\sum_{\mu\leq p}\xi_{i}^{\mu} \Big(A^\mu(t)-\frac{\kappa}{\sqrt{N}}\phi_i(t)\xi_i^\mu\Big) \\
A^\mu(t)&=& A_{e}(t) + \Omega_{\mu} +
\frac{1}{\sqrt N}\sum_{j}\phi_j(t)\xi_{j}^{\mu}
\label{eq:Amu}
\end{eqnarray}
with
$\xi^{\mu}_i=\frac{1}{2}(R^{i1}_{\mu}-R^{i2}_{\mu})$,
    $\omega^{\mu}_i=\frac{1}{2}(R^{i1}_{\mu}+R^{i2}_{\mu})$, $\Omega_{\mu}=N^{-1/2}\sum_i \omega^{\mu}_i$, and
    $\phi_i(t)=\sgn[q_i(t)]$. We have also added a perturbation field $\theta_i(t)$ to define response functions later.
Finally we introduce a spherical constraint into the model, of the type proposed in \cite{2nd Spherical}:
we replace the previous (so-called batch MG) definition $\phi_i(t)=\sgn[q_i(t)]$ by
\begin{eqnarray}
\phi_i(t)=q_i(t)/\lambda(t),~~~~~~
\lambda(t)=[N^{-1}\sum_i q_i(t)^2]^{1/2}
\end{eqnarray}
The above spherical MG version is a generalization of \cite{2nd Spherical}. It is not unique; the alternative spherical MG in \cite{1st Spherical} differs from the present formulation in at what stage and for which variables the
relevant nonlinearities are replaced by pseudo-linear laws.

\section{Generating functional analysis}

\subsection{Derivation of exact order parameter equations}

The generating functional analysis (GFA) method is based on the calculation of the
generator $\overline{Z[\bpsi]}=\overline{\bra\exp[-\rmi\sum_{it}\psi_i(t)\phi_i(t)]\ket}$, by interchanging the averages $\bra\ldots\ket$
over paths (which here, in the absence of decision noise, reduce to averaging over initial conditions) and $\overline{\cdots}$ over the disorder
(i.e. the strategies). It has become the standard tool to study MG dynamics, so we may refer to e.g. \cite{Book2} for technical details. For $N\to\infty$
and upon choosing $\theta_i(t)=\theta(t)$ for all $i$, the method leads to the following self-consistent equations for two-time correlation- and response functions $C_{tt^\prime}$ and $G_{tt^\prime}$, written in terms of averages $\bra\ldots\ket_\star$ over an effective single agent process:
\begin{eqnarray}
t\neq t^\prime:&~~~&
C_{tt^\prime}=\bra\phi(t)\phi(t^\prime)\ket_\star,~~~~~~G_{tt^\prime}=\partial \bra \phi(t)\ket_\star/\partial\theta(t^\prime)
\end{eqnarray}
A further order parameter $\lambda(t)$ is to be solved from
 $C_{tt}=1$, reflecting the spherical constraint, and causality ensures that $G_{tt^\prime}=0$ for all $t\leq t^\prime$.
The effective single agent process is defined by the following stochastic equation, with $\phi(t)=q(t)/\lambda(t)$:
\begin{eqnarray}
q(t+1)=q(t)+\theta (t) - \alpha \sum_{t^\prime\leq t}\big[\left(\one\!+\! G
\right)^{-1}\!\!-\kappa\one\big]_{tt^\prime} \phi(t^\prime)+\sqrt \alpha
\eta(t)
\label{eq:effective_agent}
\end{eqnarray}
Here $\eta(t)$ is a zero-average Gaussian noise, characterized by
 $\bra \eta(t)\eta(t^\prime)\ket=
\Sigma[A_e]_{tt^\prime}$,
\begin{eqnarray}
\Sigma[A_e]_{tt^\prime}&=&[(\one+G)^{-1}D[A_e](\one+G^{\dag})^{-1}]_{tt^\prime}
\label{eq:Sigma}
\\
D[A_e]_{tt^\prime}&=& 1+C_{tt^\prime}+2A_e(t)A_e(t^\prime)
\label{eq:D}
\end{eqnarray}
The effective Gaussian noise replaces the statistics of the original $N$
agents by evolution uncertainty for one \emph{effective} agent. We can now appreciate the
advantages of the spherical version. The effective equation (\ref{eq:effective_agent}) is linear in $\phi(t)$,
which allows us to derive fully explicit dynamical equations for the order parameters. The only nontrivial step in this derivation is
to show via integration by parts that $\bra \eta(t)\phi(t^\prime)\ket_\star=(\Sigma[A_e]G^\dag)_{tt^\prime}$.
For $\theta(t)\to 0$ the final result is
\begin{eqnarray}
\hspace*{-15mm}
\lambda(t\!+\!1)C_{t+1,t^\prime}- [\lambda(t)+\alpha\kappa]C_{tt^\prime}&=&
\alpha[(\one+G)^{-1}D[A_e](\one+G^{\dag})^{-1}G^{\dag}]_{tt^\prime}
\nonumber
\\
\hspace*{-15mm}&&
- \alpha [\left(\one + G \right)^{-1}C]_{tt^\prime}
\label{eq:Cdynamics}
\\[1mm]
\hspace*{-15mm}
\lambda(t+1)G_{t+1,t^\prime}-[\lambda(t)+\alpha\kappa]G_{tt^\prime}&=&\delta_{tt^\prime}-\alpha
\left[(\one+G)^{-1}G\right]_{tt^\prime}
\label{eq:Gdynamics}
\end{eqnarray}
By solving these equations for the kernels $\{C,G\}$, together with the condition $C_{tt}=1$ from which to extract $\lambda(t)$,
we can explore the dynamics of the original MG model for $N\to\infty$, since the physical meaning of $\{C,G\}$
is
\begin{eqnarray}
C_{tt^\prime}&=&\lim_{N\to\infty}\frac{1}{N}\sum_i \overline{\bra \phi_i(t)\phi_i(t^\prime)\ket}
\\
 G_{tt^\prime}&=&\lim_{N\to\infty}\frac{1}{N}\sum_i \overline{\partial\bra \phi_i(t)\ket/\partial\theta_i(t^\prime)}
\end{eqnarray}
The advantages of the spherical
MG are that one can derive an exact formula for the volatility (see below),
and that the explicit nature of its order parameter equations allows us to analyze the effects of the bid
perturbations  $A_e(t)$ much more effectively that in ordinary MGs.

\subsection{Bid average and fluctuations}

The statistics of the overall bids $A^\mu(t)$ in (\ref{eq:Amu})
can once more be extracted from a suitable generating functional, namely
$\overline{Z[\bvarphi]} =\overline{\bra \exp[\rmi\sqrt{2}\sum_{\mu
t}\varphi_{\mu}(t)A^{\mu}(t)]\ket}$. It generates the relevant moments of the overall bids via differentiation, e.g.
\begin{eqnarray}
\overline{\bra A^{\mu}(t) \ket}&=& -\frac{\rmi}{\sqrt
2}\lim_{\bvarphi \rightarrow \bnull}\frac{\partial \overline{Z[\bvarphi]}}{\partial
\varphi_{\mu}(t)}
\\
\overline{\bra A^{\mu}(t)A^{\nu}(t^\prime)\ket} &=&
-\frac{1}{2}\lim_{\bvarphi \rightarrow \bnull}\frac{\partial^2
\overline{Z[\bvarphi]}}{\partial\varphi_{\mu}(t)\partial\varphi_{\nu}(t^\prime)}
\end{eqnarray}
Following the familiar steps of the generating functional
analysis technique leads us back to the previous saddle-point problem, but now we obtain an additional expression for the bid moments.
We refer to \cite{Book2,DeSanctisGalla} for full details of such calculations, and limit ourselves here to giving the final result:
\begin{eqnarray}
\hspace*{-15mm}
\overline{\bra A^{\mu}(t) \ket}&=&
\sum_{t^\prime}(\one+G)^{-1}_{tt^\prime}A_{e}(t^\prime)
\label{eq:bidmoments}
\\
\hspace*{-15mm}
\overline{\bra A^\mu(t)A^\nu(t^\prime)\ket}&=& \overline{\bra A^{\mu}(t) \ket}~\overline{\bra A^{\nu}(t^\prime) \ket}+\frac{1}{2}\delta_{\mu\nu}
[(\one+G)^{-1}D_0[A_e](\one+G^{\dag})^{-1}]_{tt^\prime}
\end{eqnarray}
where $D_0[A_e]_{tt^\prime}=1+C_{tt^\prime}$ (note that $D_0[A_e]$ still depends on $A_e(t)$ via the kernel $C$).
Clearly, in the presence of finite external bid perturbations
the system is no longer guaranteed to evolve towards a state with zero-average bid statistics.
We can now define a fluctuation volatility by the following expression:
\begin{eqnarray}
\sigma_{\rm fl}^2&=& \lim_{\tau\to\infty} \frac{1}{\tau p}\sum_{t\leq \tau}\sum_\mu\Big\{ \overline{\bra[ A^\mu(t)]^2\ket}-\overline{\bra A^{\mu}(t) \ket}^{~2}\Big\}\nonumber
\\
&=& \lim_{\tau\to\infty}\frac{1}{2\tau}\sum_{t\leq \tau}[(\one+G)^{-1}D_0[A_e](\one+G^{\dag})^{-1}]_{tt}
\label{eq:volatility_formula}
\end{eqnarray}
Note, however, that in the presence of non-stationary $A_e(t)$, even in the absence of anomalous response and upon assuming self-averaging with respect to the disorder
it will generally no longer be true that $\sigma_{\rm fl}$ is identical
to the disorder-averaged conventional volatility $\sigma$ as defined by\begin{eqnarray}
 \sigma^2&=&\lim_{\tau\to\infty}\frac{1}{\tau p}\sum_{t\leq \tau}\sum_\mu [A^\mu(t)]^2
- \Big[\lim_{\tau\to\infty}\frac{1}{\tau p}\sum_{t\leq \tau}\sum_\mu A^\mu(t)\Big]^2
\label{eq:normal_volatility}
\end{eqnarray}
The difference between $\sigma$ and $\sigma_{\rm fl}$ reflects bid oscillations  which are deterministic
and therefore excluded from $\sigma_{\rm fl}$. We will derive an exact relation between the two later.

\section{Time translation invariance with constant or oscillating external bids}

 In the remainder of this paper we focus
 on the choices  $A_e(t)=\tilde{A}$ and
$A_e(t)=\tilde{A}(-1)^t$, where the asymptotic consequences of bid perturbation are most easily quantified.
Experience with previous MG versions suggests that there will be
two types of players in the stationary state: `frozen' agents, with
$q_i(t)$ growing linearly with time, and `fickle' agents, where $q_i(t)$ does not diverge with
time. In view of this we consider two types of solutions with
respect to the spherical constraint parameter $\lambda(t)$: a regime where $\lim_{t\to\infty}\lambda(t)=\lambda$ (finite), and another
regime where
 $\lambda(t)\to\infty$ as $t\to\infty$.

\subsection{General formulae}

For $A_e(t)=\tilde{A}(-1)^{\zeta t}$ with $\zeta\in\{0,1\}$ our equations (\ref{eq:Cdynamics},\ref{eq:Gdynamics}) will have
 time-translation invariant (TTI) solutions, since here (\ref{eq:D}) gives
$D[A_e]_{tt^\prime}=1+C_{tt^\prime}+2\tilde{A}^2 (-1)^{\zeta(t-t^\prime)}$.
Upon assuming $C_{tt^\prime}=C(t-t^\prime)$ and $G_{tt^\prime}=G(t-t^\prime)$, so that the same is true for $D[A_e]$ and $\Sigma[A_e]$, our equations (\ref{eq:Cdynamics},\ref{eq:Gdynamics})
then reduce for finite values of $t$ to
\begin{eqnarray}
\hspace*{-18mm}
C(t\!+\!1)- \psi_0 C(t)&=&
\alpha\psi_1
\Big\{\room
[(\one\!+G)^{-1}D[A_e](\one\!+G^{\dag})^{-1}G^{\dag}](t)
-  [\left(\one\! + G \right)^{-1}C](t)\Big\}
\nonumber \\[-1mm]
\hspace*{-18mm}&&
\label{eq:CTTI}
\\[-1mm]
\hspace*{-18mm}
G(t\!+\!1)-\psi_0 G(t)&=&
\psi_1\Big\{\delta_{t0}-\alpha
\left[(\one\!+G)^{-1}G\right](t)\Big\}
\label{eq:GTTI}
\end{eqnarray}
with
\begin{eqnarray}
\psi_0=\lim_{t\to \infty}[\lambda(t\!-\!1)\!+\!\alpha\kappa]/\lambda(t),~~~~~~\psi_1=\lim_{t\to \infty}1/\lambda(t)
\label{eq:psis}
\end{eqnarray}
If $\lambda(t)\to\lambda$ for $t\to\infty$ one has $\psi_0=(\lambda+\alpha\kappa)/\lambda$ and $\psi_1=1/\lambda$, whereas
if $\lambda(t)\to\infty$ one has $\psi_0=1$ and $\psi_1=0$ (note: since strategy valuations can diverge at most linearly with time,
the same must be true for $\lambda(t)$ so $\lim_{t\to\infty}\lambda(t-1)/\lambda(t)<1$ is ruled out).
There is, however, an important subtlety. It is not clear that in the case of diverging $\lambda(t)$ one can use equations
(\ref{eq:CTTI},\ref{eq:GTTI}) to calculate observables such as $\chi$ or $\hat{\chi}$, as this requires that the limits $\tau\to\infty$ in $\chi=\sum_{t\leq \tau}G(t)$ and $t\to\infty$ in (\ref{eq:psis}) commute. If they do not commute (as will be the case), this means that
the time it takes to evolve from initialization to a state with $(\psi_0,\psi_1)=(1,0)$ diverges, so such states will in practice not be observed.
Equations (\ref{eq:CTTI},\ref{eq:GTTI}) invite us to switch to Fourier transforms,
\begin{eqnarray}
C(t) = \int_{-\pi}^{\pi}\!\frac{\rmd\omega}{2\pi} \rme^{\rmi\omega t}\hat{C}(\omega),~~~~~~
\hat{C}(\omega)= \sum_{t=-\infty}^{\infty}\rme^{-\rmi\omega
t}C(t)
\end{eqnarray}
The Fourier transform of the kernel (\ref{eq:D}), in particluar, is seen to be
\begin{eqnarray}
\hat{D}[A_e](\omega)&=&
2\pi\delta(\omega)+\hat{C}(\omega)+4\pi\tilde{A}^2\delta(\omega-\zeta\pi)
\end{eqnarray}
One expects the relevant static observables to include  the
 integrated responses to static and oscillatory fields, viz.
$\chi=\sum_{t}G(t)=\hat{G}(0)$ and
$\hat{\chi}=\sum_{t}(-1)^tG(t)=\hat{G}(\pi)$. With these definitions we can write (\ref{eq:CTTI},\ref{eq:GTTI})
as
\begin{eqnarray}
&&
\hat{C}(\omega)\Big[
|1\!+\!\hat{G}(\omega)|^2(\rme^{\rmi\omega}-\psi_0)
+\alpha\psi_1[1+\hat{G}(-\omega)-
\hat{G}(\omega)]\Big]\nonumber
\\
&&\hspace*{40mm}=
2\pi\alpha\psi_1
\hat{G}(\omega)[\delta(\omega)+2\tilde{A}^2\delta(\omega-\zeta\pi)]
\label{eq:Cfourier}
\\
&&
(\rme^{\rmi\omega}-\psi_0)\hat{G}(\omega)=
\psi_1\Big[1-\alpha\hat{G}(\omega)/[1+\hat{G}(\omega)]\Big]
\label{eq:Gfourier}
\end{eqnarray}
We must also enforce the spherical constraint $C(0)=1$, which gives $\int\!\rmd\omega\hat{C}(\omega)=2\pi$.
The right-hand side of (\ref{eq:Cfourier}) dictates that the solution must me of the following form, where we have built in the spherical constraint, with $c_0\in[0,1]$:
\begin{eqnarray}
&&
\hat{C}(\omega)=2\pi[c_0\delta(\omega)+(1-c_0)\delta(\omega-\pi)]
\end{eqnarray}
Equations (\ref{eq:Cfourier},\ref{eq:Gfourier}) then reduce to the following set:
\begin{eqnarray}
&&
c_0[
\alpha\psi_1
+(1\!+\!\chi)^2(1\!-\!\psi_0)]=
\alpha\psi_1
\chi[1+2\tilde{A}^2\delta_{\zeta 0}]
\label{eq:stat1}
\\
&&
(1\!-\!c_0)[
\alpha\psi_1
-
(1\!+\!\hat{\chi})^2(1\!+\!\psi_0)]
=2\alpha\psi_1\hat{\chi}\tilde{A}^2\delta_{\zeta 1}
\label{eq:stat2}
\\
&&
(1-\psi_0)\chi(1+\chi)=
\psi_1(1+\chi-\alpha\chi)
\label{eq:stat3}
\\
&&
-(1+\psi_0)\hat{\chi}(1+\hat{\chi})=
\psi_1(1+\hat{\chi}-\alpha\hat{\chi})
\label{eq:stat4}
\end{eqnarray}
We can next work out the bid statistics in TTI states. Here formulas (\ref{eq:bidmoments},\ref{eq:volatility_formula})
are seen to give the following, which shows (as expected) that the effects
of stationary or oscillatory external bid perturbations are generally reflected in the bid averages:
\begin{eqnarray}
\hspace*{-15mm}
&&
\lim_{\tau\to\infty} \frac{1}{\tau}\sum_{t\leq \tau}\overline{\bra A^\mu(t)\ket}=\frac{\tilde{A}\delta_{\zeta 0}}{1+\chi},~~~~~~
\lim_{\tau\to\infty} \frac{1}{\tau}\sum_{t\leq \tau}(-1)^t\overline{\bra A^\mu(t)\ket}=\frac{\tilde{A}\delta_{\zeta 1}}{1+\hat{\chi}}
\label{eq:bidaverage_TTI}
\\
\hspace*{-15mm}
&& \sigma_{\rm fl}^2= \frac{1\!+\!c_0}{2(1+\chi)^2}+\frac{1\!-\!c_0}{2(1+\hat{\chi})^2}
\label{eq:volatility_TTI}
\end{eqnarray}

\subsection{Solution with diverging constraining force: fully frozen states}

In states where $\lambda(t)\to\infty$ for $t\to\infty$ one has $(\psi_0,\psi_1)=(1,0)$, so
equations (\ref{eq:CTTI},\ref{eq:GTTI}) reduce to $C(t)=1$ and $G(t)=0$ for all $t$, indicating a fully frozen microscopic state
with $c_0=1$.
The simplest potential solution of our remaining equations is that where we subsequently assume
the time limits in the definition of $\{\chi,\hat{\chi}\}$ to commute with those in (\ref{eq:psis}). It then
follows from (\ref{eq:stat1},\ref{eq:stat2},\ref{eq:stat3},\ref{eq:stat4}) that $\chi=\hat{\chi}=0$, and the volatility (\ref{eq:volatility_TTI}) would become
$\sigma_{\rm fl}^2= 1$. Unfortunately, the relevant limits do not commute, which follows from a more careful analysis of the effective single agent process. Although still $c_0=1$, the calculation of $\chi$ and $\hat{\chi}$ has to be done explicitly. Upon defining $Q=\lim_{t\to\infty}q(t)/t$, $\Lambda=\lim_{t\to\infty}\lambda(t)/t$, $\overline{\eta}=\lim_{t\to\infty}t^{-1}\sum_{t^\prime\leq t}\eta(t^\prime)$, and $\overline{\theta}=\lim_{t\to\infty}t^{-1}\sum_{t^\prime\leq t}\theta(t^\prime)$, and upon writing (\ref{eq:effective_agent}) in integrated form,
one finds
\begin{eqnarray}
Q&=&\frac{\overline{\theta}+\sqrt{\alpha}\overline{\eta}}{1+(\alpha/\Lambda)[(1+\chi)^{-1}-\kappa]}
\label{eq:Q}
\end{eqnarray}
This gives $Q(\overline{\eta})$, where $\overline{\eta}$ is a zero-average Gaussian random variable with $\bra\overline{\eta}^2\ket_\star=
\lim_{t\to\infty}t^{-1}\sum_{ss^\prime\leq t}\bra \eta(s)\eta(s^\prime)\ket=
(1+c_0+2\tilde{A}^2\delta_{\zeta 0})/(1+\chi)^2=2(1+\tilde{A}^2\delta_{\zeta 0})/(1+\chi)^2$.
The susceptibility now follows from $\chi=\Lambda^{-1}\partial\bra Q(\overline{\eta})\ket_\star/\partial\overline{\theta}=[\Lambda\sqrt{\alpha}]^{-1}\bra \partial Q(\overline{\eta})/\partial\overline{\eta}\ket_\star$, and $\Lambda$ follows from the spherical constraint $\lim_{t\to\infty}\bra \phi^2(t)\ket_\star=1$, which translates into $\bra Q^2(\overline{\eta})\ket_\star=\Lambda^2$. So we find the following two coupled equations, respectively,
\begin{eqnarray}
\chi&=& \frac{1}{\Lambda[1+(\alpha/\Lambda)[(1+\chi)^{-1}-\kappa]]}
\\
\Lambda&=& \frac{\sqrt{2\alpha(1+\tilde{A}^2\delta_{\zeta 0})}}{(1+\chi)[1+(\alpha/\Lambda)[(1+\chi)^{-1}-\kappa]]}
\end{eqnarray}
To calculate $\hat{\chi}$ we require an infinitesimal oscillating field $\theta(t)=(-1)^t\tilde{\theta}$, and use
\begin{eqnarray}
\hat{\chi}&=&\lim_{\tilde{\theta}\to 0}\frac{\partial}{\partial\tilde{\theta}} \lim_{\tau\to\infty}\frac{1}{\tau}\sum_{t\leq \tau}(-1)^t\bra  \frac{q(t)}{\lambda(t)}\ket_\star=0
\end{eqnarray}
by virtue of $q(t)/\lambda(t)\to Q(\overline{\eta})/\Lambda$ as $t\to\infty$.
Solving the above equations for $\chi$ and $\lambda$ and substituting the result into (\ref{eq:volatility_TTI}) gives
\begin{eqnarray}
\chi&=& \frac{1}{\sqrt{2\alpha(1+\tilde{A}^2\delta_{\zeta 0})}-1}
\\
\Lambda&=&-1-\alpha(1-\kappa)+\sqrt{\alpha}(3+2\tilde{A}^2\delta_{\zeta 0})/\sqrt{2(1\!+\!\tilde{A}^2\delta_{\zeta 0})}
\\
\sigma_{\rm fl}^2&=& \Big(1-1/\sqrt{2\alpha(1+\tilde{A}^2\delta_{\zeta 0})}\Big)^2
\end{eqnarray}
Ergodicity breaks down when $\chi\to\infty$, i.e. at
$\alpha_{c,1}(\tilde{A})=\frac{1}{2}(1+\tilde{A}^2\delta_{\zeta 0})^{-1}$. Secondly, we see that the above solution breaks down when $\lambda(t)$ no longer diverges with time, which happens when $[1+\alpha(1-\kappa)]^2=\frac{1}{2}\alpha(3+2\tilde{A}^2\delta_{\zeta 0})^2/(1\!+\!\tilde{A}^2\delta_{\zeta 0})$, i.e. at
\begin{eqnarray}
\hspace*{-15mm}
\alpha_{c,\pm}(\tilde{A},\kappa)&=&
\frac{R\!-\!2(1\!-\!\kappa)\pm \sqrt{R^2\!-\!4(1\!-\!\kappa)R}}{2(1\!-\!\kappa)^2},~~~~~~R=\frac{(3\!+\!2\tilde{A}^2\delta_{\zeta 0})^2}{2\!+\!2\tilde{A}^2\delta_{\zeta 0}}
\label{eq:alphac_finitelambda}
\end{eqnarray}
The solution $\alpha_{c,-}$ is unphysical as it obeys $\alpha_{c,-}\leq \alpha_{c,1}$, which implies that it occurs in the nonergodic regime where the above formulae are no longer valid. The solution $\alpha_{c,+}$ is relevant as it obeys $\alpha_{c,+}\geq \alpha_{c,1}$.
 Thus the regime of diverging $\lambda(t)$ is bounded on the left (as a function of $\alpha$)  by a non-ergodicity transition at $\alpha=\alpha_{c,1}(\tilde{A})$ and on the right by a transition to states with finite $\lambda(t)$ at $\alpha=\alpha_{c,+}(\tilde{A},\kappa)$.
 The transition value $\alpha_{c,+}(\tilde{A},\kappa)$ increases monotonically with increasing $\kappa$; it is minimal at $\kappa=0$, taking the value
 $\alpha_{c,+}(\tilde{A},0)=\frac{1}{2}[R\!-\!2+ \sqrt{R^2\!-\!4R}]$,
and diverges at $\kappa=1$.

\subsection{Solutions with finite constraining force}

In the alternative scenario $\lim_{t\to\infty}\lambda(t)=\lambda\in\R$ we substitute into (\ref{eq:stat1},\ref{eq:stat2},\ref{eq:stat3},\ref{eq:stat4}) the values $\psi_0=(\lambda+\alpha\kappa)/\lambda$ and $\psi_1=1/\lambda$. If we also write $\lambda$ as $\lambda=\frac{1}{2}\alpha (\gamma -\kappa)$, to simplify our equations further, we find
\begin{eqnarray}
&&
c_0=
\frac{\chi+2\chi\tilde{A}^2\delta_{\zeta 0}}{1
-\kappa(1\!+\!\chi)^2},~~~~~~
1-c_0
=\frac{2\hat{\chi}\tilde{A}^2\delta_{\zeta 1}}{1-\gamma(1\!+\!\hat{\chi})^2}
\label{eq:c0_eqns}
\\
&&
\alpha\kappa\chi^2+\chi[1+\alpha(\kappa-1)]+1=0
\label{eq:chi_eqn}
\\
&&
\alpha\gamma\hat{\chi}^2+\hat{\chi}[1+\alpha(\gamma-1)]+1=0
\label{eq:chihat_eqn}
\end{eqnarray}
Solving the last two equations for $\chi$ and $\hat{\chi}$ gives two possibilities for each:
\begin{eqnarray}
\chi_\pm=\frac{\alpha(1\!-\!\kappa)\!-\!1\pm \sqrt{[\alpha(1\!-\!\kappa)\!-\!1]^2-4\alpha\kappa}}{2\alpha\kappa}
\label{eq:chi_solns}
\\
\hat{\chi}_\pm=\frac{\alpha(1\!-\!\gamma)\!-\!1\pm \sqrt{[\alpha(1\!-\!\gamma)\!-\!1]^2-4\alpha\gamma}}{2\alpha\gamma}
\label{eq:chihat_solns}
\end{eqnarray}
It follows from the demand $\chi\in\R$ that
solutions with a finite constraining force exist only if $[\alpha(1\!-\!\kappa)\!-\!1]^2-4\alpha\kappa\geq 0$, that is, only if
\begin{eqnarray}
\alpha\leq \tilde{\alpha}_{c,-}=\frac{1}{(1\!+\!\sqrt{\kappa})^2}~~~~{\rm or}~~~~\alpha\geq \tilde{\alpha}_{c,+}=\frac{1}{(1\!-\!\sqrt{\kappa})^2}
\end{eqnarray}
Furthermore, upon investigating the limit $\kappa\to 0$ (no impact correction) in (\ref{eq:chi_solns}) we find
 $\lim_{\kappa\to 0}\chi_-=1/(\alpha-1)$ and $\lim_{\kappa\to 0}\chi_+=\infty$, so the physical solution must be
 $\chi_-$:
\begin{eqnarray}
\chi=\frac{\alpha(1\!-\!\kappa)\!-\!1- \sqrt{[\alpha(1\!-\!\kappa)\!-\!1]^2-4\alpha\kappa}}{2\alpha\kappa}
\label{eq:chi_soln}
\end{eqnarray}
which is finite for all $\kappa>0$, and positive for $\alpha>1/(1-\kappa)$.
To find the static order parameter $c_0$ we need only combine (\ref{eq:chi_soln}) with the first equation of
(\ref{eq:c0_eqns}). The remaining two equations, viz. (\ref{eq:chihat_solns}) and the second equation of (\ref{eq:chi_soln}), serve only
to determine the quantities $\hat{\chi}$ and $\gamma$. Substituting (\ref{eq:chi_soln}) into (\ref{eq:c0_eqns}) gives
\begin{eqnarray}
c_0&=&
\frac{2\alpha (1+2\tilde{A}^2\delta_{\zeta 0})[\alpha(1\!-\!\kappa)\!-\!1- \sqrt{[\alpha(1\!-\!\kappa)\!-\!1]^2-4\alpha\kappa}]}
{4\alpha^2\kappa-[\alpha(1\!+\!\kappa)\!-\!1- \sqrt{[\alpha(1\!-\!\kappa)\!-\!1]^2-4\alpha\kappa}]^2}
\label{eq:c_soln}
\end{eqnarray}
We extract from this that $c_0=-\alpha-\alpha^2(1\!+\!2\kappa)+\order(\alpha^3)$ for small $\alpha$, and that $c_0\downarrow -\infty$ for $\alpha\uparrow\tilde{\alpha}_{c,-}$. Hence the solution (\ref{eq:chi_soln},\ref{eq:c_soln}) is unphysical in the regime $\alpha\in[0,\tilde{\alpha}_{c,-}]$.
This leaves the regime $\alpha>\tilde{\alpha}_{c,+}$. Here we are sure that $\chi>0$ (since $1/(1-\kappa)<\tilde{\alpha}_{c,+}$), and we find
$c_0=1/\alpha(1\!-\!\kappa)+2\kappa/\alpha^2(1\!-\!\kappa)^2+\order(\alpha^{-3})$ for $\alpha\to\infty$, and that $c_0\uparrow\infty$ for $\alpha\downarrow\tilde{\alpha}_{c,+}$. Hence the solution (\ref{eq:chi_soln},\ref{eq:c_soln}) is physical in the regime $\alpha\in[\tilde{\alpha}_{c,2},\infty)$,
where $\tilde{\alpha}_{c,2}(A_e,\kappa)\geq 1/(1\!-\!\sqrt{\kappa})^2$ is defined by the condition $c_0=1$. This last critical line is most easily derived indirectly, by first using (\ref{eq:c0_eqns}) to write it as a condition on $\chi$, giving
\begin{eqnarray}
\chi_{c,2}=\frac{\sqrt{(1+2\tilde{A}^2\delta_{\zeta 0})^2+8\kappa(1+\tilde{A}^2\delta_{\zeta 0})}-1-2\tilde{A}^2\delta_{\zeta 0}}{2\kappa}-1
\end{eqnarray}
This is then inserted into (\ref{eq:chi_eqn}) and the result is easily solved for $\alpha$, giving
\begin{eqnarray}
&&
\tilde{\alpha}_{c,2}(A_e,\kappa)= \frac{2(\Xi\!-\!1\!-\!2\tilde{A}^2\delta_{\zeta 0})}
{(\Xi\!-\!1\!-\!2\tilde{A}^2\delta_{\zeta 0}\!-\!2\kappa)
(3\!+\!2\tilde{A}^2\delta_{\zeta 0}\!-\!\Xi)
}
\label{eq:tilde_alpha}
\\
&&
\Xi=\sqrt{(1\!+\!2\tilde{A}^2\delta_{\zeta 0})^2\!+\!8\kappa(1\!+\!\tilde{A}^2\delta_{\zeta 0})}
\end{eqnarray}
After (tedious) reworking of this result, using the definitions of $\Xi$ and $R$ and the identity $\Xi=[2(1\!+\!\tilde{A}^2\delta_{\zeta 0})(R\!-\!4(1\!-\!\kappa))]^{1/2}$ to connect the various short-hands, one finds that this critical line (\ref{eq:tilde_alpha}) marking $c_0=1$ with finite $\lambda(t)$ is identical to the
previous critical line (\ref{eq:alphac_finitelambda}). For $\tilde{A}=0$ we recover the corresponding formula of \cite{2nd Spherical}:
\begin{eqnarray}
\tilde{\alpha}_{c,2}(0,\kappa)= \frac{4\kappa+5+3\sqrt{1+8\kappa}}{
4(1-\kappa)^2}
\end{eqnarray}
Let us finally turn to the susceptibility $\hat{\chi}$, which we need to calculate the volatility (\ref{eq:volatility_TTI}), and which is to be solved together with $\gamma$ from (\ref{eq:chihat_eqn}) and the second equation in
(\ref{eq:c0_eqns}).
One finds upon eliminating $\gamma$ that
  \begin{eqnarray}
  \hat{\chi}_\pm &=& \frac{-1}{1\pm \sqrt{\alpha[1+2\tilde{A}^2\delta_{\zeta 1}/(1\!-\!c_0)]}}
  \label{eq:chi_hat_final}
  \end{eqnarray}
  Since the present solution regime always has $\alpha>1$, we see that the two distinct solutions $\hat{\chi}_\pm$ represent in-phase and
  out-of-phase
  responses of the market to the oscillating external bid input, and that neither will be able to diverge.
The volatility (\ref{eq:volatility_TTI}) now follows upon
substituting the
result (\ref{eq:chi_hat_final}), together with (\ref{eq:chi_soln}) and (\ref{eq:c_soln}).

\section{Phase diagrams}

We can now summarize the picture obtained by analyzing the time-translation invariant states in terms of a phase diagram, where the control parameters are $\alpha$, $\kappa$, and $\tilde{A}$.
The three phases of the system are the following:\\[-2mm]
\begin{eqnarray}
\hspace*{-15mm}
\alpha\in(\alpha_{c,2},\infty):&~~~{\rm oscillating~phase~(O)}, & ~~~ c_0<1,~\chi<\infty,~\hat{\chi}\neq 0,~\lambda(t)<\infty
\nonumber \\
\hspace*{-15mm}
\alpha\in(\alpha_{c,1},\alpha_{c,2}):&~~~{\rm frozen~phase~(F)},      & ~~~ c_0=1,~\chi<\infty,~\hat{\chi}=0,~\lambda(t)\to\! \infty
\nonumber \\
\hspace*{-15mm}
\alpha\in[0,\alpha_{c,1}):&~~~{\rm anomalous~phase~(A)}, & ~~~ c_0=1,~\chi=\infty
\nonumber\\[-2mm] \hspace*{-15mm}\nonumber
\end{eqnarray}
The precise
dependence of the transition lines $\alpha_{c,1}$ and $\alpha_{c,2}$ on the two remaining control parameters $(\tilde{A},\kappa)$
is furthermore controlled strongly by whether the external bid perturbation is static ($\zeta=0$) or oscillatory ($\zeta=1$). In the former case one finds
\begin{eqnarray}
\hspace*{-20mm}
\alpha_{c,1}(\tilde{A})&=&\frac{1}{2}(1+\tilde{A}^2)^{-1}
\\
\hspace*{-20mm}
\alpha_{c,2}(\tilde{A},\kappa)&=&
\frac{R(\tilde{A})\!-\!2(1\!-\!\kappa)+\! \sqrt{R^2\!(\tilde{A})\!-\!4(1\!-\!\kappa)R(\tilde{A})}}{2(1\!-\!\kappa)^2},~~~~~R(a)=\frac{(3\!+\!2a^2)^2}{2\!+\!2a^2}
\end{eqnarray}
Here the frozen (F) phase will consistently grow with increasing values of $\tilde{A}$, due to both its left boundary $\alpha_{c,1}(\tilde{A})$ decreasing with $\tilde{A}$ and its right boundary $\alpha_{c,2}(\tilde{A},\kappa)$ increasing with $\tilde{A}$.
This is easily understood: a static external bid perturbation diminishes the impact of the internal
bid contributed by the agents, and makes it easier for the agents to find a suitable strategy that will land them in the minority group.
At $\kappa=0$ (no self-impact correction) the transition point $\alpha_{c,2}$ simplifies to $\alpha_{c,2}(\tilde{A},0)=2(1+\tilde{A}^2)=1/\alpha_{c,1}(\tilde{A})$.
For oscillatory perturbations $A_e(t)=\tilde{A}(-1)^t$, in contrast, the situation is very different: here the relevant transition lines are strictly independent of the amplitude $\tilde{A}$, and are
obtained by evaluating the above formulae at
$\alpha_{c,1}(0)$ and $\alpha_{c,2}(0,\kappa)$, viz.
 \begin{eqnarray}
\alpha_{c,1}=\frac{1}{2},~~~~~~
\alpha_{c,2}(\kappa)=\frac{5+4\kappa+3\sqrt{1+8\kappa}}{4(1-\kappa)^2}
\end{eqnarray}
 However, although in the case of oscillatory perturbation the phase diagram is independent of $\tilde{A}$, in the oscillating (O) phase there will be a significant dependence on $\tilde{A}$ of the volatility.
In all cases we find that the effect of self-impact correction is to strengthen the frozen (F) phase, with the oscillating phase vanishing
altogether for full impact correction (i.e. for $\kappa=1$).

\begin{figure}[t]
\vspace*{2mm} \hspace*{-1mm} \setlength{\unitlength}{0.58mm}
\begin{picture}(250,100)

  \put(0,0){\includegraphics[height=100\unitlength,width=140\unitlength]{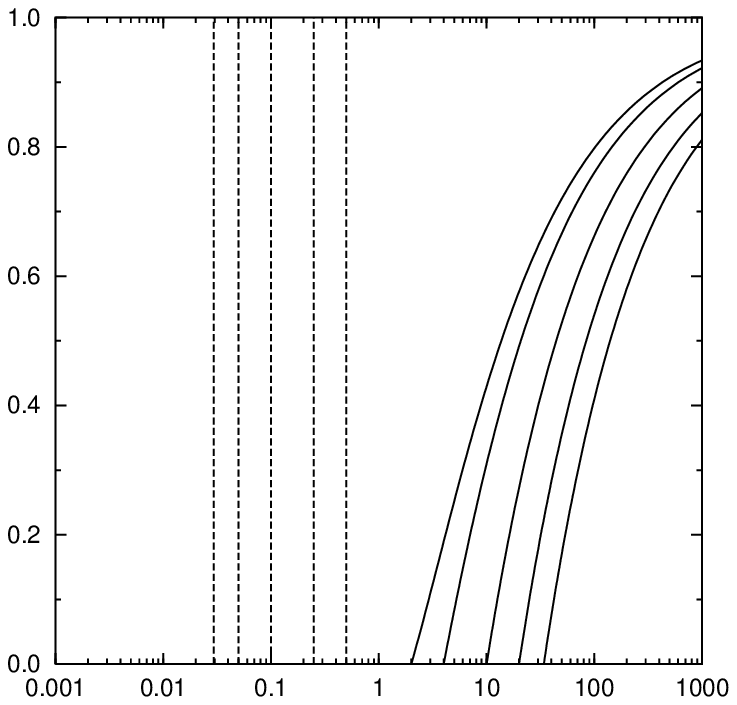}}
  \put(-3,55){\large\here{$\kappa$}}  \put(54,-12){\large$\alpha$}
  \put(18,50){\large A}  \put(65,80){\large F}  \put(90,20){\large O}

  \put(120,0){\includegraphics[height=100\unitlength,width=140\unitlength]{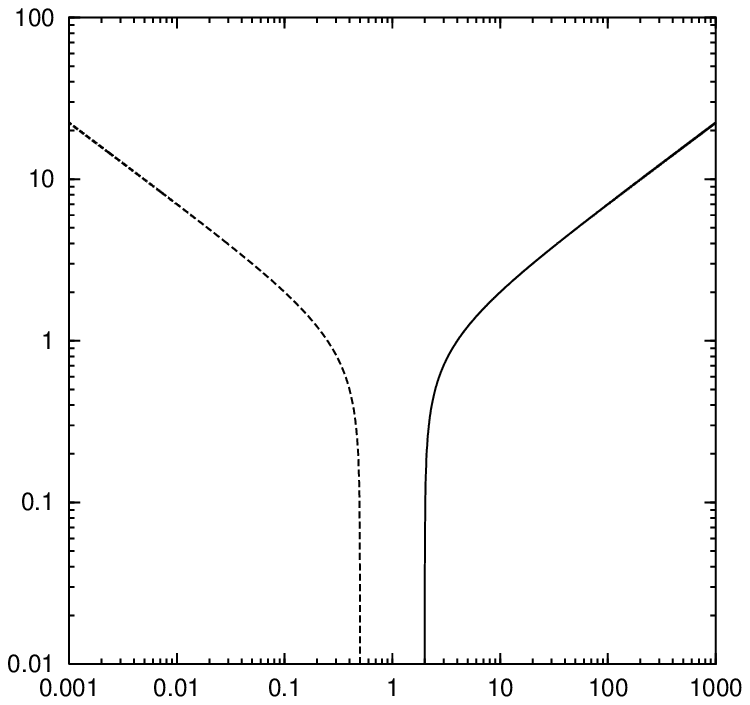}}
  \put(119,55){\large\here{$\tilde{A}$}}  \put(174,-12){\large$\alpha$}
  \put(150,30){\large A}  \put(175,80){\large F}  \put(200,30){\large O}

\end{picture}
 \vspace*{5mm}
\caption{Phase diagrams for static external bid perturbations, $A_e(t)=\tilde{A}$, exhibiting an oscillating phase (O), a frozen phase (F), and an anomalous phase (A). The latter is characterized by $\chi=\infty$.
Dashed lines: the F$\to$A transition $\alpha_{c,1}$. Solid lines: the F$\to$O transition $\alpha_{c,2}$.
Left diagram: transitions in the $(\alpha,\kappa)$ plane. Here we show the lines for $\tilde{A}=0,1,2,3,4$ (left to right in the case of F$\to$O, right to left in the  case of  F$\to$A). Right diagram: transitions in the $(\alpha,\tilde{A})$ plane for $\kappa=0$ (no self-impact correction), where $\alpha_{c,2}(\tilde{A},0)=1/\alpha_{c,1}(\tilde{A})$. For static external bid perturbation, increasing $\tilde{A}$ always increases the F phase.
}\label{fig:phasediagrams_static}
\end{figure}

\begin{figure}[h]
\vspace*{5mm} \hspace*{-30mm} \setlength{\unitlength}{0.58mm}
\begin{picture}(250,100)

  \put(120,0){\includegraphics[height=100\unitlength,width=140\unitlength]{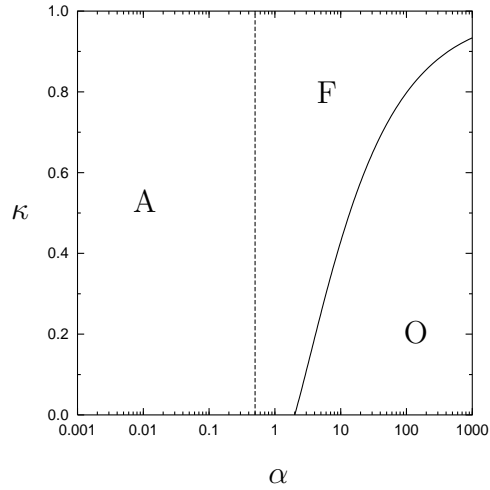}}
  \put(117,50){\large\here{$\kappa$}}  \put(174,-12){\large$\alpha$}
  \put(143,50){\large A}  \put(185,75){\large F}  \put(205,20){\large O}

\end{picture}
 \vspace*{8mm}
\caption{Phase diagrams for oscillating
external bid perturbations, $A_e(t)=\tilde{A}(-1)^t$, with the same definitions of phases and transitions as in the previous figures.
For oscillating bid perturbations the phase diagram is independent of the perturbation amplitude $\tilde{A}$.
}\label{fig:phasediagram_oscill}
\end{figure}

It follows from expressions (\ref{eq:bidaverage_TTI}) that throughout the phases F and O the system is not able to compensate fully
for stationary external bid perturbations; since $\chi$ is finite, the average bid will be nonzero for any nonzero perturbation
amplitude $\tilde{A}$. Only at the transition line $\alpha_{c,1}(\tilde{A})$, where nonergodicity sets in and $\chi$ diverges, does the bid average
vanish. This is the situation that was also encountered in ordinary MGs \cite{DeSanctisGalla}. In our present spherical MG, however, we are able to inspect also the impact of oscillating external bid perturbations, and find that with such perturbations the situation is worse: since $\hat{\chi}$ remains finite everywhere, we will always retain
an oscillating (and predictable) nonzero overall bid average, even if we enter the nonergodic regime.

\section{Tests against numerical simulations}

\begin{figure}[t]
\vspace*{3mm} \hspace*{-10mm} \setlength{\unitlength}{0.47mm}
\begin{picture}(250,200)

  \put(0,5){\includegraphics[height=90\unitlength,width=140\unitlength]{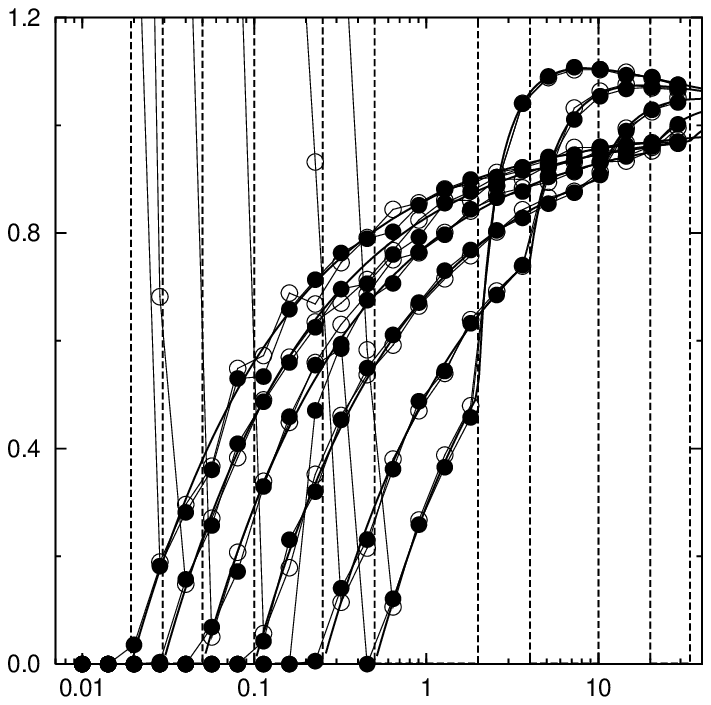}}   \put(54,-7){\large$\alpha$}
  \put(100,5){\includegraphics[height=90\unitlength,width=140\unitlength]{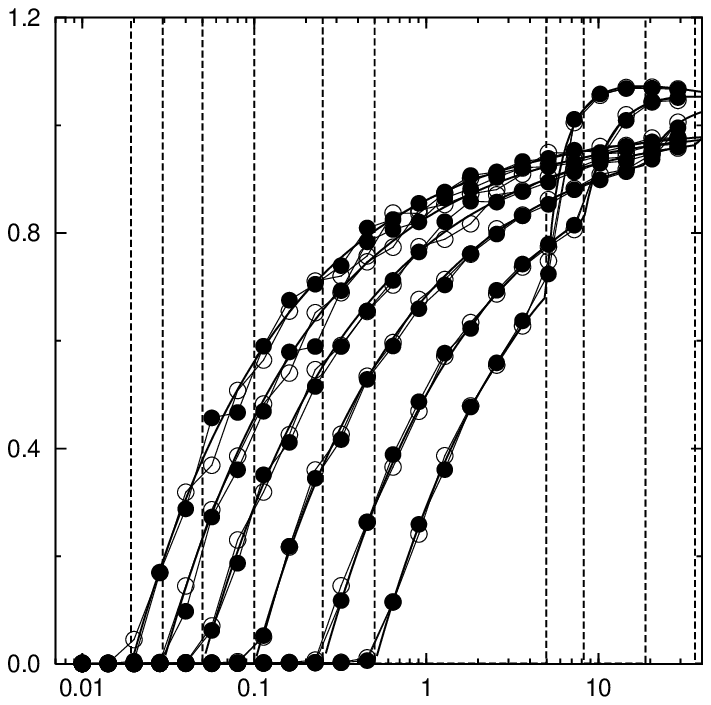}} \put(154,-7){\large$\alpha$}
  \put(200,5){\includegraphics[height=90\unitlength,width=140\unitlength]{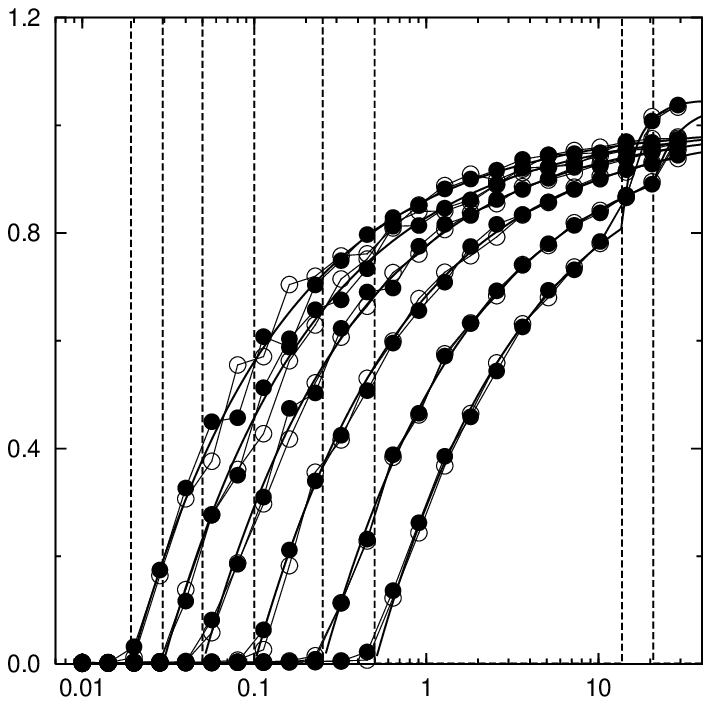}} \put(254,-7){\large$\alpha$}

 \put(0,100){\includegraphics[height=90\unitlength,width=140\unitlength]{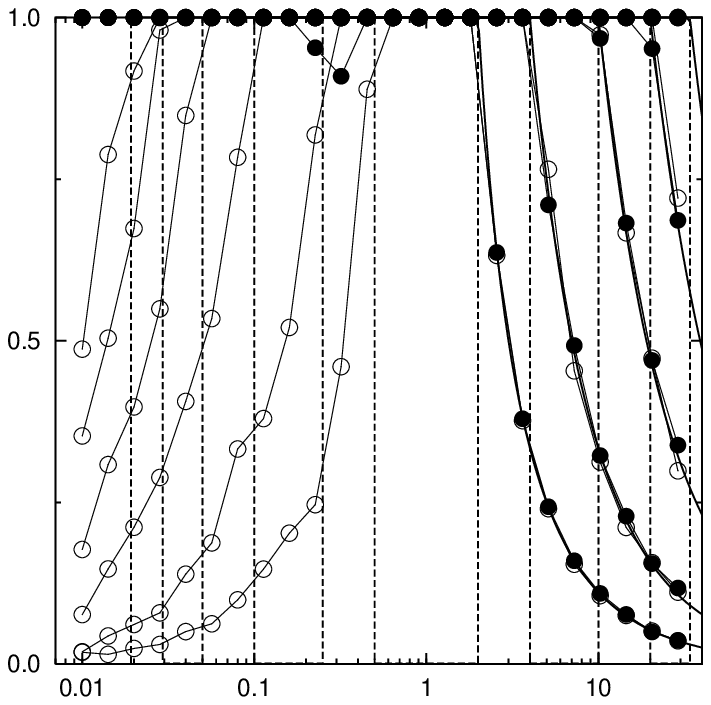}}
 \put(100,100){\includegraphics[height=90\unitlength,width=140\unitlength]{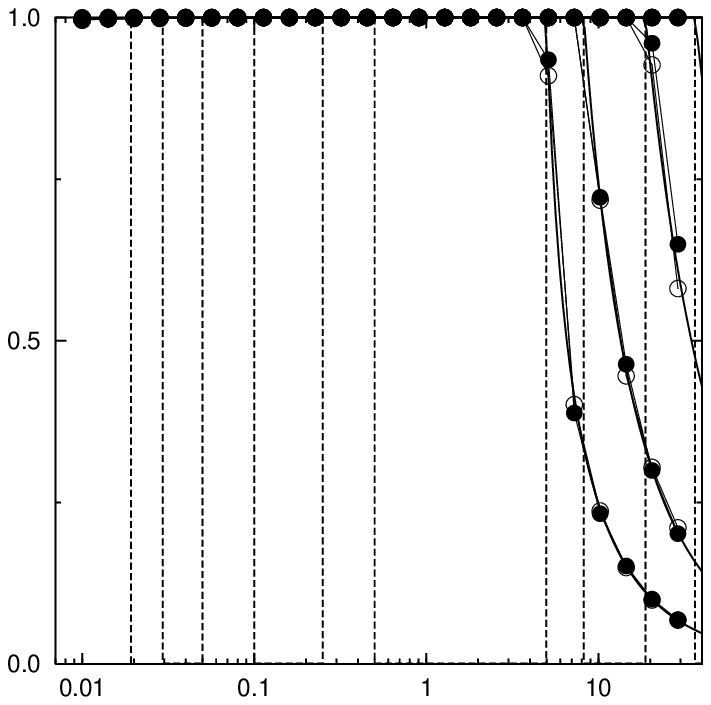}}
 \put(200,100){\includegraphics[height=90\unitlength,width=140\unitlength]{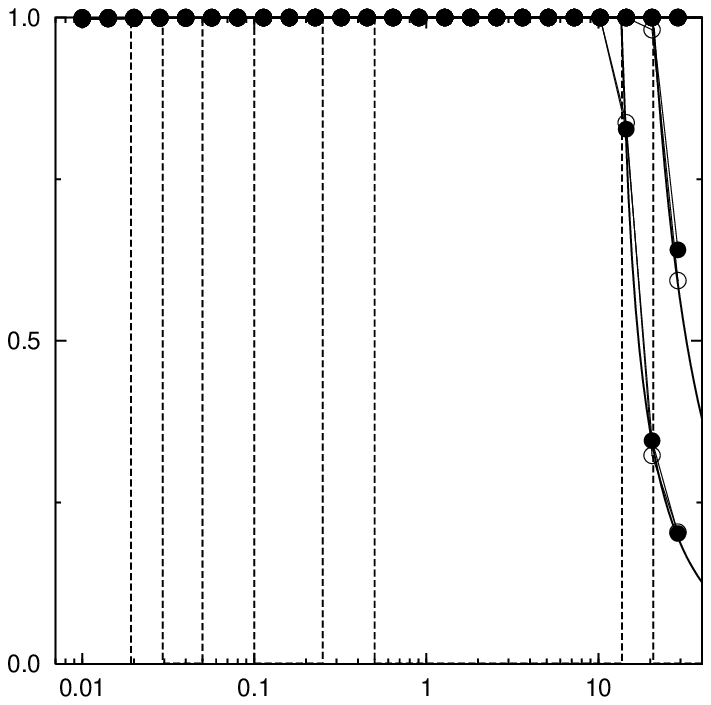}}

\put(-7,50) {\large $\sigma$} \put(-7,150){\large $c_0$}
\put(42,194){\large $\kappa=0$} \put(140,194){\large $\kappa=0.25$} \put(240,194){\large $\kappa=0.5$}
\end{picture}
 \vspace*{5mm}
\caption{Macroscopic observables for static  external bid contributions $A_e(t)=\tilde{A}$.
Top row: frozen correlations $c_0$ for $\kappa=0$ (left), $\kappa=0.25$ (middle) and $\kappa=0.5$ (right); with in each picture
the values as measured for $\tilde{A}\in\{0,1,2,3,4,5\}$.
Bottom row: volatility $\sigma$ for $\kappa=0$ (left), $\kappa=0.25$ (middle) and $\kappa=0.5$ (right); with in each picture
the values as measured for $\tilde{A}\in\{0,1,2,3,4,5\}$. In all figures the connected markers indicate simulation results (full circles: biased initial conditions; open circles: unbiased initial conditions), with solid lines showing the corresponding theoretical predictions. The vertical dashed lines mark the transition values $\alpha_{c,1}(\tilde{A})$ (left, transition from the nonergodic regime with finite spherical constraint force regime for small $\alpha$ to  the frozen regime) and $\alpha_{c,2}(\tilde{A},\kappa)$ (right, transition from the frozen regime with infinite spherical constraint force to the large $\alpha$ ergodic regime with finite spherical constraint force).
}\label{fig:nonoscillating}
\end{figure}

\begin{figure}[t]
\vspace*{3mm} \hspace*{-10mm} \setlength{\unitlength}{0.48mm}
\begin{picture}(250,200)

  \put(0,5){\includegraphics[height=90\unitlength,width=140\unitlength]{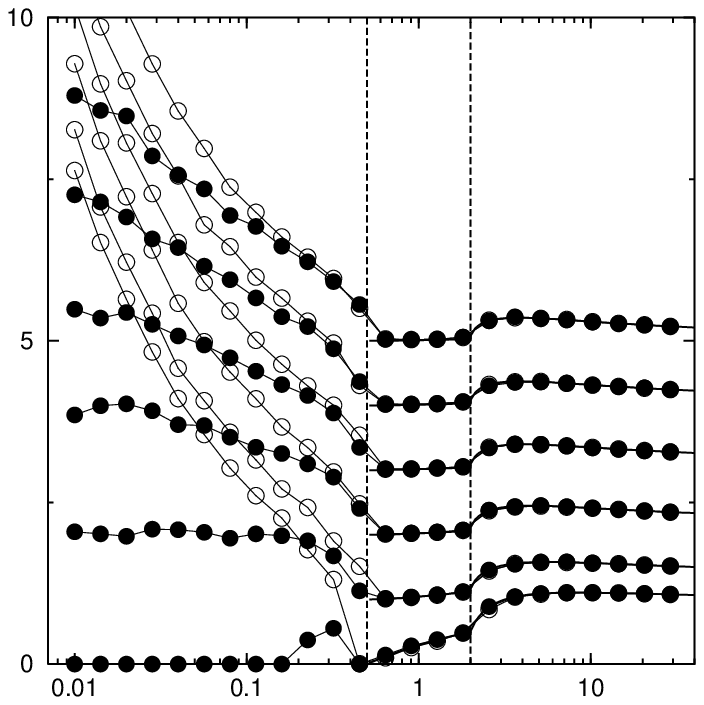}}   \put(54,-7){\large$\alpha$}
  \put(100,5){\includegraphics[height=90\unitlength,width=140\unitlength]{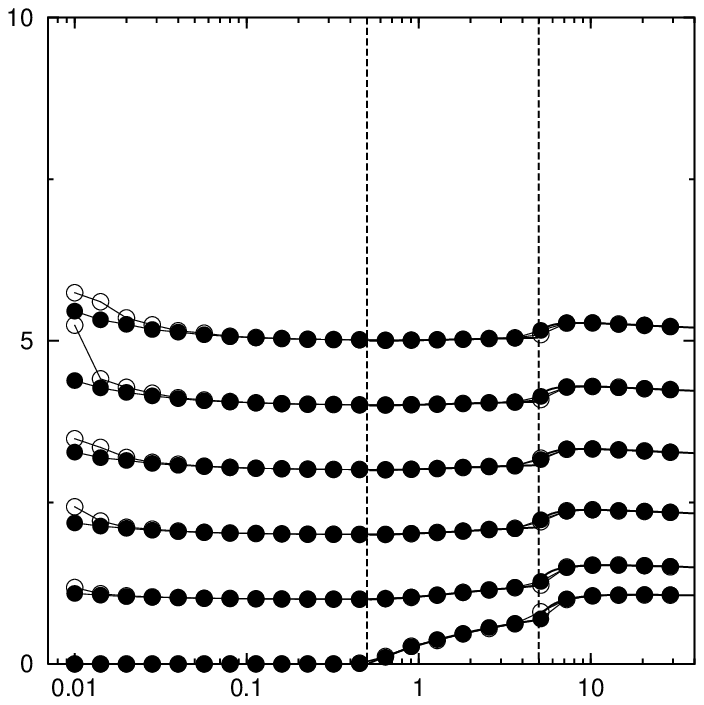}} \put(154,-7){\large$\alpha$}
  \put(200,5){\includegraphics[height=90\unitlength,width=140\unitlength]{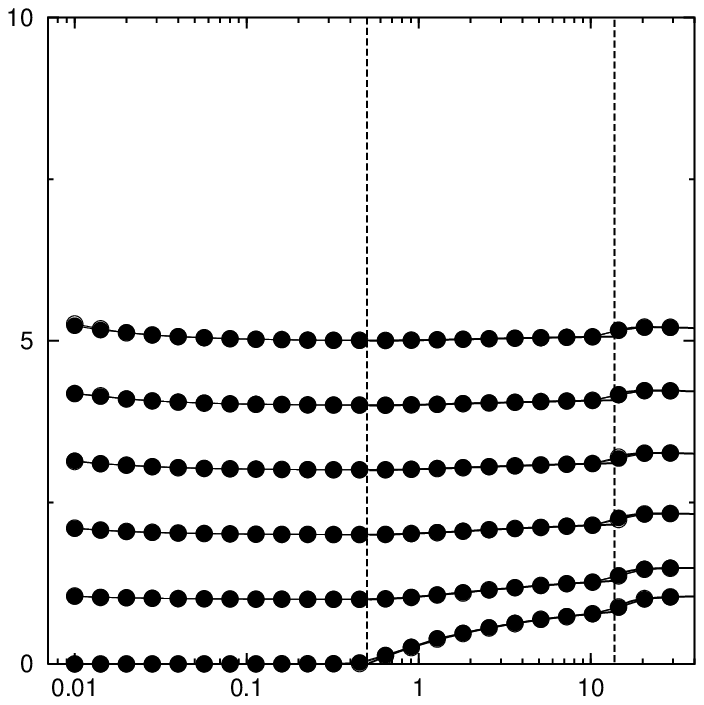}} \put(254,-7){\large$\alpha$}

 \put(0,100){\includegraphics[height=90\unitlength,width=140\unitlength]{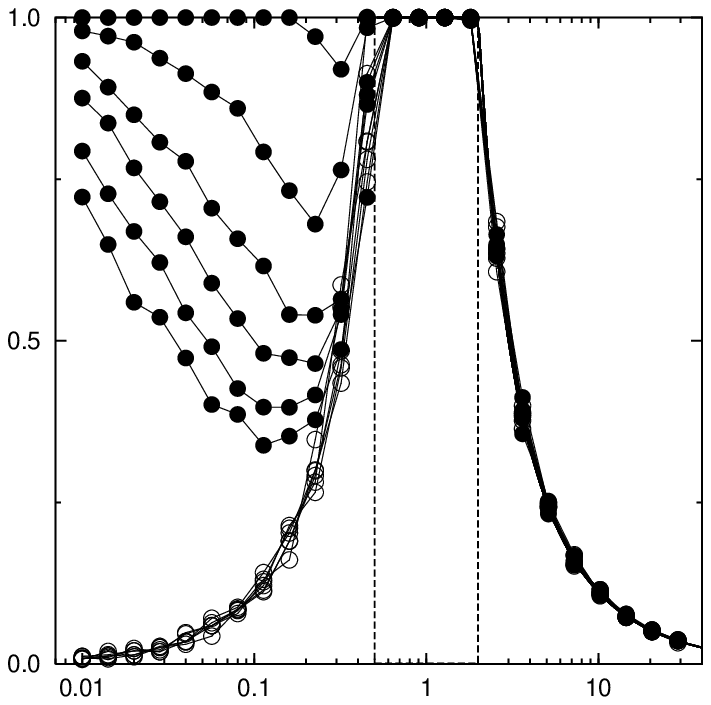}}
 \put(100,100){\includegraphics[height=90\unitlength,width=140\unitlength]{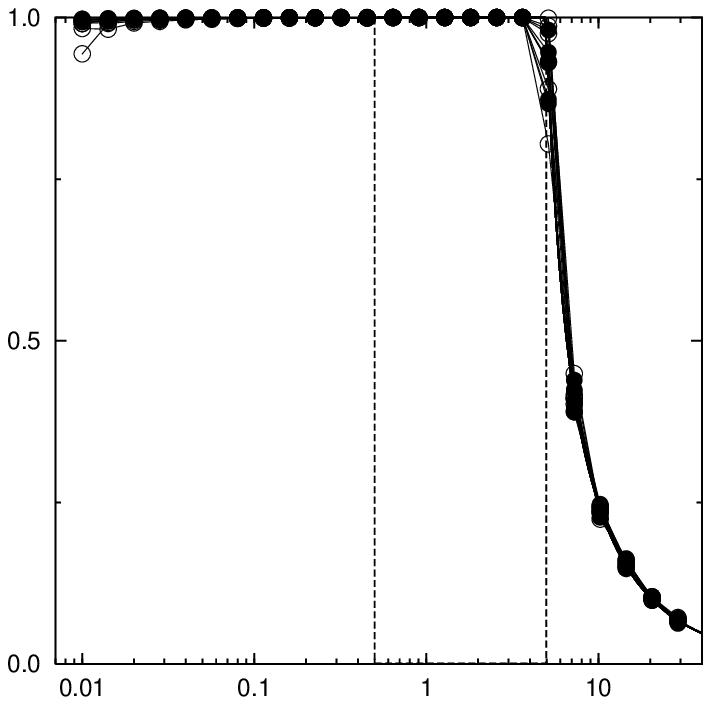}}
 \put(200,100){\includegraphics[height=90\unitlength,width=140\unitlength]{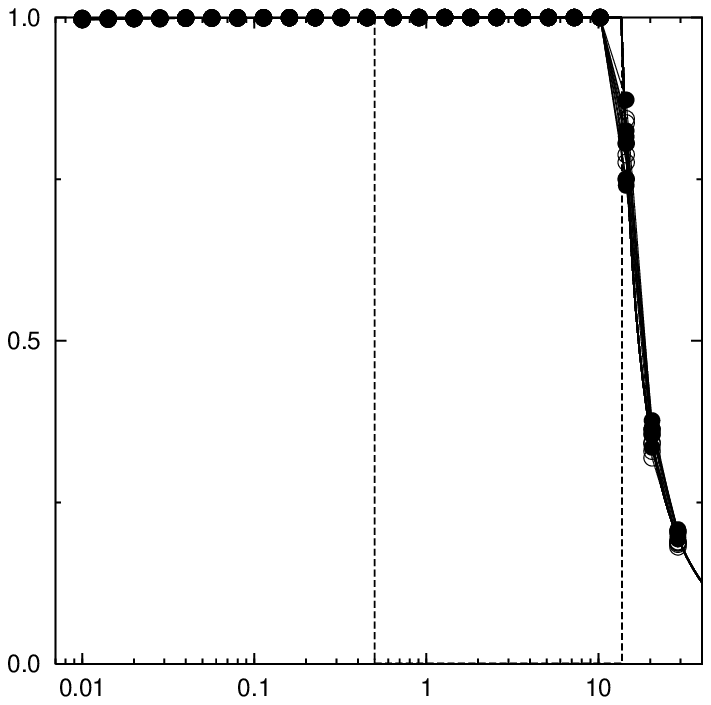}}

\put(-7,50) {\large $\sigma$} \put(-7,150){\large $c_0$}
\put(42,194){\large $\kappa=0$} \put(140,194){\large $\kappa=0.25$} \put(240,194){\large $\kappa=0.5$}
\end{picture}
 \vspace*{5mm}
\caption{Macroscopic observables for oscillating external bid contributions $A_e(t)=\tilde{A}(-1)^t$.
Top row: frozen correlations $c_0$ for $\kappa=0$ (left), $\kappa=0.25$ (middle) and $\kappa=0.5$ (right); with in each picture
the values as measured for $\tilde{A}\in\{0,1,2,3,4,5\}$.
Bottom row: volatility $\sigma$ for $\kappa=0$ (left), $\kappa=0.25$ (middle) and $\kappa=0.5$ (right); with in each picture
the values as measured for $\tilde{A}\in\{0,1,2,3,4,5\}$. Note the different vertical volatility scales compared to the previous figure. In all figures the connected markers indicate simulation results (full circles: biased initial conditions; open circles: unbiased initial conditions), with thick solid lines showing the corresponding theoretical predictions. The vertical dashed lines mark the transition values $\alpha_{c,1}=\frac{1}{2}$ (left, transition from the nonergodic regime with finite spherical constraint force regime for small $\alpha$ to  the frozen regime) and $\alpha_{c,2}(\kappa)$ (right, transition from the frozen regime with infinite spherical constraint force to the large $\alpha$ ergodic regime with finite spherical constraint force).
}\label{fig:oscillating}
\end{figure}

\begin{figure}[ht]
\vspace*{5mm} \hspace*{10mm} \setlength{\unitlength}{0.53mm}
\begin{picture}(250,100)

  \put(0,5){\includegraphics[height=90\unitlength,width=140\unitlength]{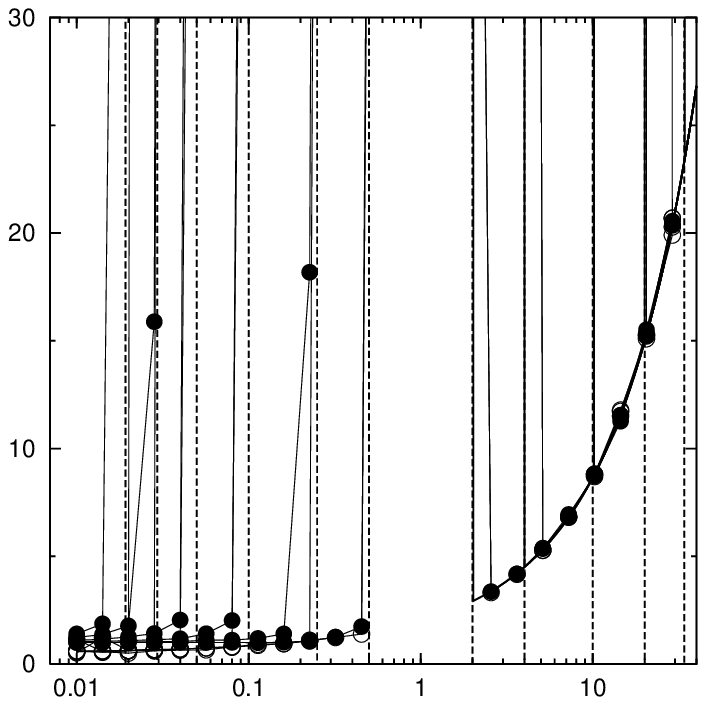}}   \put(54,-7){\large$\alpha$}
  \put(120,5){\includegraphics[height=90\unitlength,width=140\unitlength]{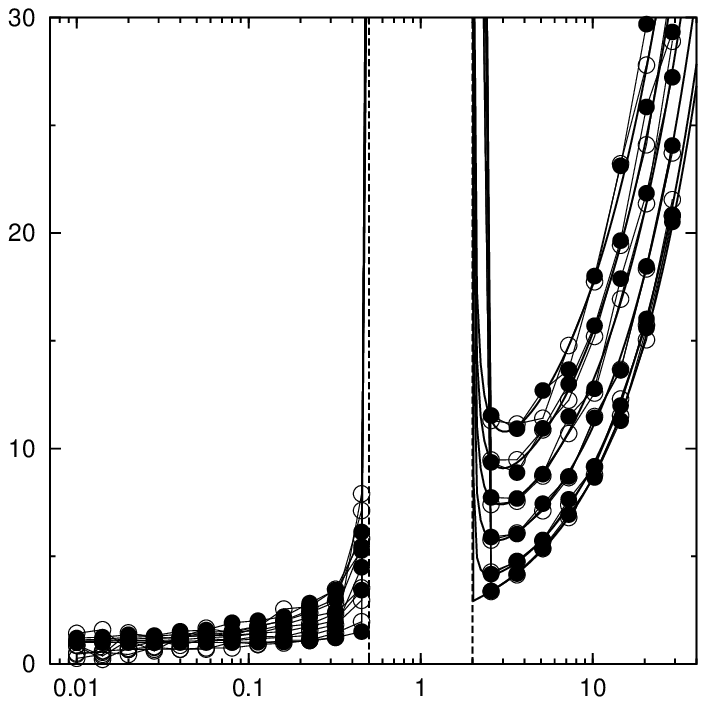}} \put(174,-7){\large$\alpha$}

\put(-7,50) {\large $\lambda$} \put(113,50) {\large $\lambda$}
\put(42,98){\large $A_e(t)=\tilde{A}$} \put(150,98){\large $A_e(t)=\tilde{A}(-1)^t$}
\end{picture}
 \vspace*{0mm}
\caption{Examples of measured values (connected markers) versus predicted values (thick solid lines) of the spherical constraint
force $\lambda$, as measured for $\tilde{A}\in\{0,1,2,3,4,5\}$ with $\kappa=0$. The vertical dashed lines mark the transition values $\alpha_{c,1}$ (left, transition from the nonergodic regime with finite spherical constraint force regime for small $\alpha$ to  the frozen regime) and $\alpha_{c,2}$ (right, transition from the frozen regime with infinite spherical constraint force to the large $\alpha$ ergodic regime with finite spherical constraint force). The force $\lambda$ is is indeed seen to diverge at the predicted values of $\alpha$.
}\label{fig:lambda}
\end{figure}

In this section we test our predictions regarding the values of the long-time order parameters and the locations of phase transition lines, for stationary and oscillating external bid contributions, respectively,
against numerical simulations. All simulations were carried out with systems of size $N=3000$, for both unbiased ($q_i(0)=\pm 10^{-4}$)  and biased
($q_i(0)=\pm 1$) random initializations. Observables were always measured over a duration of 2000 batch iterations, following an equilibration stage of 1000 batch iterations. When comparing theoretical predictions to simulation measurements it became immediately clear that of the two possible saddle-points
in (\ref{eq:chi_hat_final}) the physical one is $\hat{\chi}_+$, the in-phase solution with high volatility; henceforth all theoretical predictions will correspond to this saddle-point.

The volatility $\sigma$ that is measured in simulations is the conventional one given in (\ref{eq:normal_volatility}), for which we still have to derive an analytical expression. Upon combining definition (\ref{eq:volatility_formula}) of $\sigma_{\rm fl}$, for which we have derived prediction (\ref{eq:volatility_TTI}), with
(\ref{eq:normal_volatility}) (following any given initial state, so that the brackets $\bra \ldots\ket$ are irrelevant, and assuming self-averaging of both volatilities over the disorder realization for $N\to\infty$) we see that
\begin{eqnarray}
 \sigma^2&=&\sigma_{\rm fl}^2
 +\lim_{\tau,p\to\infty}\Big\{ \frac{1}{\tau p}\sum_{t\leq \tau}\sum_\mu \overline{ A^{\mu}(t) }^{~2}
- \frac{1}{\tau^2 p^2}\sum_{tt^\prime\leq \tau}\sum_{\mu\nu} \overline{ A^\mu(t)A^\nu(t^\prime)}
\Big\}
\nonumber
\\
&=& \sigma_{\rm fl}^2
 +\lim_{\tau\to\infty}\Big\{ \frac{1}{\tau}\sum_{t\leq \tau}\Big[\sum_{t^\prime}(\one+G)^{-1}_{tt^\prime}A_{e}(t^\prime)\Big]^{2}
 \nonumber
 \\
 &&
 -\Big[\frac{1}{\tau}\sum_{t\leq \tau}\sum_{t^\prime}(\one+G)^{-1}_{tt^\prime}A_{e}(t^\prime)\Big]^2\Big\}
\end{eqnarray}
In time-translation invariant stationary states with $A_e(t)\in\{\tilde{A},\tilde{A}(-1)^t\}$ this becomes
\begin{eqnarray}
 \sigma^2&=& \sigma_{\rm fl}^2+\sum_{ss^\prime}(\one+G)^{-1}(s)(\one+G)^{-1}(s^\prime)
 \nonumber
 \\
 &&\times
\lim_{\tau\to\infty}\Big\{ \frac{1}{\tau}\sum_{t\leq \tau}A_{e}(t\!-\!s)A_e(t\!-\!s^\prime)
 -\frac{1}{\tau^2}\sum_{tt^\prime\leq \tau}A_{e}(t\!-\!s)A_e(t^\prime\!-\!s^\prime)\Big\}\nonumber
 \\
 &=&
 \sigma_{\rm fl}^2~+~\left\{
 \begin{array}{lll} \tilde{A}^2/(1+\hat{\chi})^2 &~{\rm if}~& A_e(t)=\tilde{A}(-1)^t\\
 0 &~{\rm if}~& A_e(t)=\tilde{A}
 \end{array}
 \right.
\end{eqnarray}
The results are shown in figures \ref{fig:nonoscillating}, \ref{fig:oscillating}, and \ref{fig:lambda}. One observes excellent agreement between theory and numerical experiments, both in terms of the values of the observables and the locations of the O$\to$F and F$\to$A transitions. In particular, we find confirmation of the somewhat surprising prediction that, although in terms of market stability the oscillating external bid term
is a disaster for the market (the market always responds in-phase to the oscillation, i.e. making the deviations from market efficiency consistently worse), the locations of the phase transition lines remain always identical to what they were in the absence of oscillatory external disruption (i.e. for $\tilde{A}=0$).  The effects of impact correction are relatively minor,  limited to enlarging the frozen phase F, to dampening the fluctuations, and to reducing the non-ergodicity effects in the anomalous phase A.

\section{Discussion}

In this paper we have studied analytically the response of spherical minority games (MG) to external time-dependent market disruption of the overall market bid, which is envisaged to model the effects of either market regulators or other socio-political or natural events outside the market
that impact directly on the asset demand and hence the asset price. The advantages of the spherical formulation of the game are that one can derive fully
explicit (and exact) macroscopic dynamical equations and work out the market's response rigorously and effectively, even if one includes self-impact correction (where agents correct their actions for their own impact on the market), and that the volatility can be calculated exactly.
We have focused specifically on evaluating and comparing the long-time consequences of two types of external bid terms: stationary ones, $A_e(t)=\tilde{A}$, and oscillating ones, $A_e(t)=\tilde{A}(-1)^t$.
These two cases give rise to different phase diagrams and different behaviour of the market volatility.
The presence of an oscillating external bid term is disasterous for the efficiency of the market: the market's response is
always to oscillate
{\em in phase} with the perturbation, making the situation worse and increasing significantly the volatility.  Yet, rather surprisingly, the phase diagram in this case is not at all affected by the imposed external bid oscillation. All this is more or less the opposite of how the market responds to stationary external bid terms, where the phase diagram is strongly affected by the amplitude of the bid perturbation, but the fluctuations
remain largely unaffected.
All theoretical predictions have been tested against numerical simulations, and found to be supported perfectly well.
 Our present study emphasizes the need for further development of those versions of the generating functional analysis formalism that
 include the stochastic overall bid dynamics explicitly, as in  \cite{RealHistories,InnerProduct}, so that one can also tackle the MG market's response  to events in its environment for non-spherical (i.e. more realistic) versions, where the route to explicit macroscopic equations that proved fruitful in this paper is no longer available.
\\[3mm]
{\bf Acknowledgements}
\\[3mm]
It is our pleasure to thank Tobias Galla and Isaac P\'{e}rez Castillo for stimulating discussions.

\end{document}